\documentclass[12pt]{article}

\usepackage{epsfig}
\def\be{\begin{equation}}
\def\ee{\end{equation}}
\def\ba{\begin{eqnarray}}
\def\ea{\end{eqnarray}}

\def\Zop{\bbbz}

\def\I{{\cal I}}

\def\bbbz {{\sf Z\!\!Z}}

\def\half{{1\over 2}}
\def\Tr{{\rm Tr}}
\def\ie{{\it i.e.}}

\begin{document}

\thispagestyle{empty}
\def\thefootnote{\fnsymbol{footnote}}
\begin{flushright}
 DAMTP-2000-52 \\
 hep-th/0005211\\
 \end{flushright}
\vskip 0.5cm

\begin{center}\LARGE
{\bf The Type I D-instanton and}\\
{\bf its M-theory Origin}
\end{center}
\vskip 1.0cm
\begin{center}
{\centerline{  Tathagata Dasgupta, Matthias R. Gaberdiel and Michael
 B. Green\footnote{{\tt T.Dasgupta,
 M.R.Gaberdiel, M.B.Green@damtp.cam.ac.uk}}}}

\vskip 0.5 cm
{\it Department of Applied Mathematics and Theoretical Physics
\\ Wilberforce Road, Cambridge, CB3 0WA, U.K.}
\end{center}

\vskip 2.0cm

%\begin{center}
%February 2000
%\end{center}
%\vskip 1.0cm

\begin{abstract}
The tree-level amplitude for the scattering of two gauge particles
constrained to move  on  the two distinct boundaries of
eleven-dimensional space-time in the Ho\v{r}ava--Witten
formulation of M-theory is constructed. At low momenta this
reproduces the corresponding tree-level scattering amplitude of
the $E_8\times E_8$ heterotic string theory.  After
compactification to nine dimensions on a large circle with a
suitable Wilson line to break the symmetry to $SO(16) \times
SO(16)$ this amplitude is used to describe the scattering of two
massive $SO(16)$ spinor states -- one from  each factor of the
unbroken symmetry group. The amplitude contains a component that
is associated with the exchange of a Kaluza--Klein charge between
the boundaries,  which is interpreted as the exchange of a
D-particle between orientifold planes in the Type IA theory. This
is related by T-duality to the effect of a non-BPS D-instanton in
the Type I theory which is only invariant under those elements of
$O(16)\times SO(16)$ that are in $SO(16) \times SO(16)$.

\end{abstract}

\vfill
\setcounter{footnote}{0}
\def\thefootnote{\arabic{footnote}}
\newpage

\section{Introduction}
\setcounter{equation}{0}

Over the past two years there has been a great deal of progress in the
study of non-BPS solitonic states in string theory \cite{Senrev}.
Among these are the stable non-BPS D-branes of certain orbifold
theories \cite{Sen2,BG2,Sen6,BG3}, and of the Type I (and Type IA)
theories \cite{Sen4,Sen5,Frau,BGH,DS}. All of these stable non-BPS
states can be elegantly characterized in terms of K theory
\cite{WittenK}.

Most of the above non-BPS D-branes can be obtained by suitable
projections from the corresponding unstable non-BPS states of the Type
II theories \cite{Sen7}. These unstable non-BPS D$p$-branes occur for
the complementary values of the (stable) BPS branes, {\it i.e.} they
have odd $p$ in Type IIA and even $p$ in Type IIB. However, in the
case of Type I there exists also a $\Zop_2$ D-instanton
(D$(-1)$-brane), which has the same value of $p$ as the corresponding
Type IIB D$(-1)$-brane. (The same is also the case for the Type I
D$7$-brane.) The presence of the Type I D-instanton is associated with
the fact that the actual gauge group of Type I is $SO(32)$ rather than
the gauge group of the perturbative Type I theory which is $O(32)$.

In this paper we will determine some explicit effects of the Type I
D-instanton by studying a simple scattering process in M-theory that
can be identified, via familiar dualities, with a Type I string theory
process. We will start, in section 2, by considering dynamics in the
eleven-dimensional geometry considered by Ho\v{r}ava and Witten, in
which the eleven-dimensional M-theory space-time is taken to be
$M^{10}\times R$, where $M^{10}$ is ten-dimensional Minkowski space
and $R$ is the interval $0\le x^{11} \le d$
\cite{HW1,HW2}. The tree-level contribution to the scattering of two
$E_8$ gauge particles that are confined to distinct boundaries of the
eleven-dimensional space-time will be evaluated in section 3, as
illustrated in figure 1. Since the external particles are localized on
distinct boundaries the exchanged bulk states are singlets under both
the $E_8$ gauge groups. We will verify that in the low-momentum limit
this amplitude reproduces the same expression as that obtained in the
$E_{8\,L}\times E_{8\,R}$ heterotic string (which will be referred to
as the $HE$ theory) in the same limit, where the subscripts $L$ and
$R$ refer to the two boundaries (left and right) at $x^{11}=0$ and
$x^{11}= d$.  This is true independent of the separation of the
boundaries. 

In section 4 we will consider the theory compactified on a circle,
$S^1$, in the $x^9$ direction with the symmetry broken to
$SO(16)_L \times SO(16)_R$. We will be particularly interested in the
scattering of massive spinor states localized on the two
nine-dimensional boundaries that are modes of the massless
ten-dimensional states with Kaluza--Klein charge $\pm 1/2$.  The
four-point function for these states follows very simply from the
ten-dimensional expression.  We will also describe the interpretation
of this amplitude in terms of the Type IA theory in which the
background is that of the Type IIA theory on the $\Zop_2$ orbifold
of a circle.  In this description the scattering spinor states are
D-particles ($D_S$) or  anti D-particles ($\bar D_S$) stuck to the two
orientifold  planes. The scattering amplitudes that will be considered
are `tree-level' interactions.  Processes of the form $D_S+ D_S \to
D_S+ D_S$, $D_S + \bar D_S \to D_S + \bar D_S$ are described by the
exchange of closed strings (states of zero Kaluza--Klein charge)
between the separated orientifold planes \footnote{Here the notation
$L+R \to L'+R'$ denotes a scattering process in which $L$ and $R$ are
incoming states on the left and right orientifolds, respectively,
while the corresponding outgoing states are $L'$ and $R'$.}. At low
energies the former process is BPS, and the amplitude vanishes in the
zero velocity limit. In the impact parameter description, the
resulting force is proportional to $v^4$ which is in agreement with
the string theory calculation that can be performed following
\cite{bachas}. The second process is non-BPS, and the amplitude is
singular as $v\rightarrow 0$, reflecting the instability of the system
to decay into a non-BPS D1-brane.

The process  $D_S+ \bar D_S \to \bar D_S + D_S$ is described by the
exchange of a $D0$-brane (the state of Kaluza--Klein charge $1$)
between the separated orientifold planes. The effect of a massive
particle exchange generates an exponentially small amplitude. After a
T-duality in the $x^{11}$ direction this is reinterpreted as the
effect of the Type I D-instanton. This is analogous to the
description of the Type IIB D-instanton as the dual of the world-line
of a Type IIA D-particle \cite{greengut} which also has an origin in
eleven-dimensional supergravity \cite{ggv}. As in that case, the
supergravity Feynman diagram calculation automatically takes into
account the appropriate quantum measure, including fermionic zero
modes.  These modes are
interpreted in terms of massless string states in the Type IA
picture. As expected, the D-instanton  contribution respects
$SO(16)_L\times SO(16)_R$ but is not invariant under
$O(16)_L\times SO(16)_R$ or $SO(16)_L\times O(16)_R$, as will be
described in some detail in section 4.2. Finally, the discussion in
section 5 contains some speculations about the situation in which the
four particles in the scattering amplitude are located on one of the
two boundaries in the nine-dimensional theory.

\section{Feynman diagrams in the Ho\v{r}ava-Witten geometry}

We are interested in calculating the tree-level scattering amplitude
illustrated in figure 1, in which the propagator joins the two
boundaries which are separated by a distance $d$. The
scattering particles with momenta $k_r$ ($r=1,\cdots,4$) are the
states confined to the ten-dimensional boundaries which are massless
$E_8$ gauge particles in the unbroken theory. Later we will consider
compactification on a circle, in which case the gauge group
may be broken and the scattering particles may carry non-zero
Kaluza--Klein charge and be massive states. When the external
particles are bosons (as will be the case in this paper) the
propagating intermediate state in the figure may be either a graviton,
$h_{\mu\nu}$, or a third-rank antisymmetric tensor potential,
$C^{(3)}_{\mu\nu\rho}$, and the full amplitude is obtained by summing
over both these contributions (where the eleven-dimensional indices
span the range  $\mu,\nu,\rho = 0,1,\cdots, 9,11$).\footnote{Recently
a similar calculation was undertaken by Krause \cite{Krause} but
the  three-form exchange was omitted so
 his result is therefore not complete.}

\begin{figure}[htb]
\epsfysize=8cm
\centerline{\epsffile{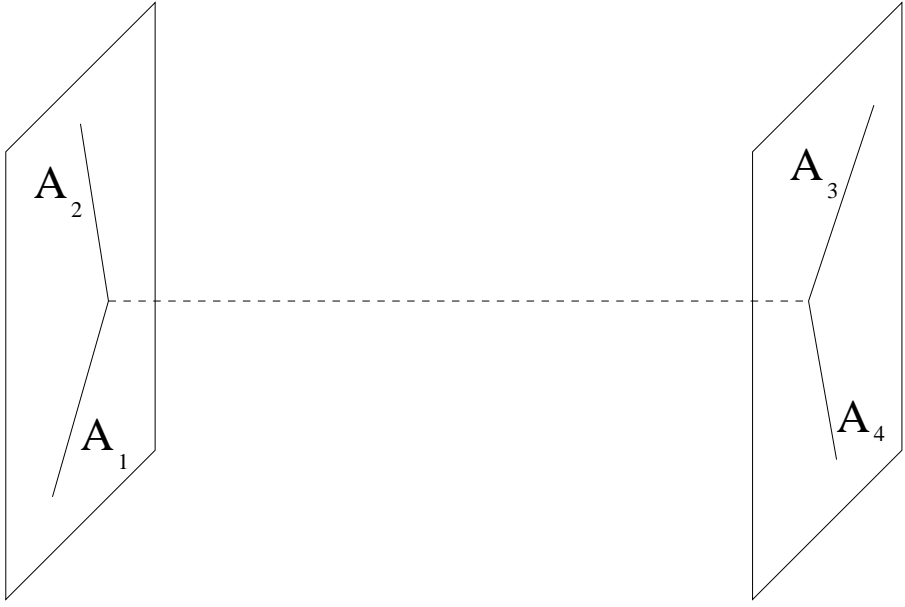}}
\caption{Tree scattering diagram of four gauge particles}
\end{figure}

The Ho\v{r}ava--Witten geometry may be thought of as a $\Zop_2$
orbifold, where the generator of the orbifold group acts on the
eleventh dimension, $x^{11}$, as $x^{11}\mapsto - x^{11}$, as well
as acting on the three-form field $C$ as $C\mapsto -C$. In the resulting
theory the momentum in the eleventh dimension, $p_{11}$, is not
conserved, whilst it is still conserved in the tangential ten
dimensions. Furthermore, the boundary conditions on the bulk fields
require that $h_{M\, 11}=0=C_{MNP}$ on the boundaries $x^{11}=0$
and $x^{11}=d$, where $M,N,P =0,1,\dots,9$ are the
ten-dimensional indices.\footnote{We are here following the `upstairs'
formalism of \cite{HW2}, in which the  fields span the covering space
$R^{1,9} \times S^1$ and the orbifold conditions are imposed by
modding out by the action of $\Zop_2$.}

We will consider the simple example of the propagator for a scalar
field in this background by first considering M-theory compactified
on  a circle with background geometry $M_{10} \times S^1$, where $S^1$
is a circle of radius  $d/\pi = R_{11} l_p$ (where $R_{11}$ is the
dimensionless radius and $l_p$ is  the eleven-dimensional Planck
length). The momentum-space scalar propagator is
\be
\tilde G(p,p_{11}) = {1\over p^2 + p_{11}^2}\,,
\label{gtilddef}
\ee
where $p_{11} = m (R_{11}l_p)^{-1} = \pi m d^{-1}$
and $p^M$ ($M=0,1,\dots, 9$) is the arbitrary
momentum in the ten dimensions tangential to the boundaries. The
scalar propagator evaluated between two points, $x^{11}$ and $y^{11}$,
on the  circle is therefore given by the Fourier sum,
\be
G(p^M; x^{11}-y^{11}) = {1 \over 2d}
\sum_{m=-\infty}^\infty e^{i{\pi m \over d} (x^{11}-y^{11})}
{1\over p^2 +{\pi^2 m^2\over d^2}}\,.
\label{mixprop}
\ee
We are really interested in defining the propagator on the $\Zop_2$
orbifold of the circle. If we choose the scalar field to be symmetric
under the action of the $\Zop_2$ then the propagator must be invariant
under $x^{11}\mapsto -x^{11}$ as well as
$x^{11}\mapsto x^{11} + 2d$ (with similar conditions on
$y^{11}$). The propagator satisfying these conditions is then given by
\be
{\cal G}(p^M; x^{11},y^{11}) =
G(p^M; x^{11}-y^{11}) + G(p^M;x^{11}+y^{11})\,.
\label{gdef}
\ee

Since the external particles are constrained to the planes $x^{11}=0$
and $x^{11}=d$, the propagator that enters in the scattering
amplitude is obtained by substituting (\ref{mixprop}) into
(\ref{gdef}) and setting $x^{11}=0, \, y^{11}= d$, which
gives,
\ba
{\cal G} (p^M; 0, d) & = & {1 \over d}
\sum_{m=-\infty}^\infty {(-1)^m \over p^2 + {\pi^2 m^2\over d^2}}
= {1\over d} \sum_{m=-\infty}^\infty (-1)^m
\int_0^\infty d\sigma ~e^{-\sigma (p^2+\frac{\pi^2 m^2} {d^2})}
\nonumber \\
&=&
\frac{1}{\sqrt\pi}\sum_{n =-\infty}^\infty
\int_{0}^\infty d\sigma
~\sigma^{-\frac{1}{2}}
e^{-\sigma p^2-\frac{d^2}{\sigma}(n+\frac{1}{2})^2}\,,
\label{schw1}
\ea
where the last line follows after performing a Poisson resummation
that replaces  the integer Kaluza--Klein charge $m$ by the integer
$2n+1$ that labels the winding number of the propagator around the
$x^{11}$ direction. (In particular, the sum in (\ref{schw1}) only
contains contributions with odd winding number; this reflects the fact
that the states in the propagator run from $x^{11}=0$ to
$x^{11}=d$.) The $\sigma$ integral is standard (the saddle
point approximation is exact) and leads to the result
\be
{\cal G} (p^M; 0, d) =
\frac{1}{\sqrt{p^2}}\sum_{n=-\infty}^\infty
e^{-|2n+1| \sqrt{p^2} d}
= \frac{1}{\sqrt{p^2}}\frac{1}{\sinh(\sqrt{p^2} d)}\,.
\label{finalprop1}
\ee
In deriving (\ref{finalprop1}) we have assumed that $p^2$ is positive
(so that the integral in (\ref{schw1}) converges). However it is clear
from the definition of ${\cal G} (p^M; 0, d)$ in
(\ref{schw1}) that ${\cal G}$ is a meromorphic function of $p^2$;
since the right-hand-side of (\ref{finalprop1}) is also meromorphic,
the final result will therefore hold in general (\ie\ not only for
positive $p^2$). We will see later how the terms in the sum over $n$
correspond to the contributions of D-instantons to the Type I theory
in nine dimensions.

In the amplitude calculations that follow, the vertices will involve
momentum factors. In the case of the exchange of the three-form
potential this will result in a term in which a factor of $p_{11}^2$
is inserted into the numerator of the propagator, so that the sum in
(\ref{schw1}) is  replaced by
\ba
\hat{\cal G}(p^M; 0, d) &=&
{1\over d} \sum_{m=-\infty}^\infty
      \frac{(-1)^m\, p_{11}^2}{p^2+p_{11}^2}
= - {1\over d} \sum_{m\in \Zop} (-1)^m\int_{0}^{\infty}
d\sigma e^{-\sigma p^2} \frac{d}{d\sigma}
e^{-\frac{\pi^2 m^2}{d^2}\sigma}
\nonumber \\
&=&  - \frac{1}{\sqrt\pi}\sum_{n\in \Zop}
\int_0^{\infty}d\sigma~e^{-\sigma p^2}
\frac{d}{d\sigma}\bigg(\sigma^{-\frac{1}{2}}
e^{-\frac{d^2}{\sigma} (n+\frac{1}{2})^2}\bigg)\nonumber \\
&=& - \sqrt{p^2} \frac{1}{ \sinh( \sqrt{p^2} d)} \,,
\label{schw2}
\ea
where a Poisson resummation has been performed in the second step, and
we have integrated by parts and used (\ref{finalprop1}) to obtain
the final line.

The propagators that enter into the tree diagrams of interest
are those of the metric and the three-form potential. Writing the
metric in non-compact eleven-dimensional space-time as
$g_{\mu\nu} = \eta_{\mu\nu} + h_{\mu\nu}$, where
$\eta_{\mu\nu}$ is the eleven-dimensional background Minkowski metric,
the $\langle h_{\mu_1\nu_1} h_{\mu_2\nu_2}\rangle$  propagator (in the
de Donder gauge) is given by
\be
\tilde G_{\mu_1\nu_1;\mu_2\nu_2}(p^M,p_{11})
= \kappa^2 \Bigg(\eta_{\mu_1\mu_2}\eta_{\nu_1\nu_2}
+ \eta_{\mu_1\nu_2} \eta_{\nu_1\mu_2}
- \frac{2}{9}\eta_{\mu_1\nu_1}\eta_{\mu_2\nu_2}\Bigg)\,
{1\over p^2 +p_{11}^2}
\,.
\label{hprop}
\ee
Similarly, writing
$C^{(3)}_{\mu\nu\rho}=c^{(3)}_{\mu\nu\rho}+\hat C^{(3)}_{\mu\nu\rho}$,
where $c^{(3)}_{\mu\nu\rho}=0$ is the background antisymmetric
potential, the
$\langle \hat C^{(3)}_{\mu_1\nu_1\rho_1}
     \hat  C^{(3)}_{\mu_2\nu_2\rho_2} \rangle$
propagator is given (in Feynman gauge) by
\be
\tilde G_{\mu_1 \nu_1 \rho_1; \mu_2 \nu_2 \rho_2}(p^M,p_{11}) =
\frac{\kappa^2}{(3!)^2} \eta_{\mu_1[\mu_2}\eta_{|\nu_1|\nu_2}
\eta_{|\rho_1|\rho_2]}\,{1\over p^2 +p_{11}^2}   \,.
\label{cprop}
\ee

The vertices that couple these propagators to the gauge particles
in the boundaries are determined by the  supergravity action in the
Ho\v{r}ava--Witten geometry which is given in \cite{HW2} and
reproduced in appendix~A.
The tree-level amplitudes involve the cubic vertices coupling a pair
of gauge particles to a graviton or an antisymmetric potential. The
graviton coupling ${A}{A}h$ is given by
\be
S_{YM}^{{A}{A}h} =   \frac{1}{(4\pi)^{5/3}\kappa^{4/3}}
\int_{R^{1,9}} d^{10}x~
\Big(\partial_M{A}^a_N\partial_P{A}^a_Qh_{RS}\Big)
\Big(\half\eta^{M[Q}\eta^{P]N}\eta^{RS} +
\eta^{M[P}\eta^{Q]S}\eta^{RN}+
\eta^{M[R}\eta^{Q]N}\eta^{PS}\Big).
\label{aahvert}
\ee
The three-form coupling ${A}{A} \hat C$ is extracted from the $G^2$
term in the bulk action. This interaction arises because the Bianchi
identity for the bulk field strength $G_{MNP\,11}$ receives a boundary
contribution in a manner that is determined by requiring local
supersymmetry as discussed in \cite{HW2},
\be
G_{MNP\,11} = 4!\partial_{[M}\hat C_{NP\, 11]}
+ \frac{\kappa^{2/3}}{\sqrt 2(4\pi)^{5/3}}
\left(\delta(x^{11}) + \delta(x^{11} - d) \right)
\omega_{MNP} \,,
\label{Ghw}
\ee
where $\omega$ is the Chern-Simons three-form defined by
\be
\omega_{MNP} = 2~{\rm Tr} \big( A_M \partial_{[N}A_{P]}
+ \frac{1}{3} A_M[A_N , A_P] + \mbox{cyclic perms.} \big)\,.
\label{cs}
\ee
The other components of $G$ are eliminated on the boundary since the
three-form field is anti-symmetric under $x^{11}\mapsto -x^{11}$. The
${A}{A}C$ interaction comes from the term linear in $\omega_{MNP}$
in $G_{MNP\, 11}^2$ and is given by
\be
S^{{A}{A}C} =   \frac{\sqrt{2}}{(4\pi)^{5/3}\kappa^{4/3}}
\int_{R^{1,9} \times S^1}
d^{11}x \left(\delta(x^{11}) + \delta(x^{11} - d) \right)
\partial_{[M}\hat C_{NP11]} \,
{\rm Tr} ({A}^M\partial^{[N}{A}^{P]}) \,.
\label{aacvert}
\ee

\section{ Scattering  of boundary gauge particles}
\setcounter{equation}{0}

We will now consider the amplitude illustrated in fig.~1 that
describes tree-level elastic scattering of an incoming gauge particle
localized on one of the boundaries with a gauge particle localized on
the other boundary. Since the  exchanged bulk fields do not carry
gauge quantum numbers the diagram has an overall group theory factor
${\rm Tr}_L(T_1T_2){\rm Tr}_R(T_3T_4)$, where ${\rm Tr}_L$ denotes the
trace over the 248 components of the adjoint representation of the
$E_{8\,L}$ gauge group on one boundary while ${\rm Tr}_R$
denotes the trace over $E_{8\,R}$ on the other. We will verify that
this amplitude reduces in the limit of small momenta to the same
expression as that of the tree-level $E_{8\,L} \times E_{8\,R}$
heterotic string with coupling constant related to the radius
$R_{11}$ in the usual manner. We will only consider the case in which
all the external states are gauge bosons, in which case the exchanged
states are the three-form potential and the graviton.

The momentum of the scattering states is restricted to the ten
flat directions parallel to the boundaries, and the Mandelstam
variables $s,t,u$ are defined in terms of the ten-dimensional momenta
$k_1,k_2,k_3,k_4$ by
\be
s = -(k_1 + k_2)^2 \,, \qquad
t = -(k_1 + k_4)^2 \,, \qquad
u = -(k_1 + k_3)^2 \,,
\ee
with $s+t+u=0$, which follows from momentum conservation,
$\sum_r k_r=0$,  and the mass-shell conditions, $k_r^2=0$, for the
external physical states. These vector states have polarization
vectors $\epsilon_r^M$ which satisfy the transversality conditions
$\epsilon_r \cdot k_r=0$.

The non-standard aspect of the tree diagrams illustrated in fig.~1 is
that the two vertices are constrained to lie in the boundaries at
$x^{11}=0$ and $x^{11}= d = \pi R_{11} l_p$. This means that momentum
is not conserved in the eleventh direction and we must use propagators
of the form ${\cal G}(p;0,d)$, suitably generalized to account for
the tensor indices on the propagating  fields as in (\ref{hprop}) and
(\ref{cprop}).

The expressions for the tree amplitude due to the graviton and the
three-form exchange are given in appendix~B by $A^{(h)}_4$ (\ref{b2}) and
$A^{(C)}_4$ (\ref{b3}).   The complete
tree-level contribution is the sum of these two amplitudes, and it is
given by
\be\label{totall}
A_4 = {\kappa^{-2/3}\over 4 (4\pi)^{10/3} } \Tr (T_1 T_2) \Tr(T_3 T_4)
\frac{1}{\sqrt{-s}}
{1 \over \sinh(\sqrt{-s}d)} t_8 F^4 \,,
\ee
where
\ba t_8 F^4 &=& \Bigl\{
- 2 u t (\epsilon_1 \cdot \epsilon_2) (\epsilon_3 \cdot \epsilon_4)
- 2 s t (\epsilon_1 \cdot \epsilon_3) (\epsilon_2 \cdot \epsilon_4)
- 2 s u (\epsilon_1 \cdot \epsilon_4) (\epsilon_2 \cdot \epsilon_3)
\nonumber \\
& & \qquad
+(\epsilon_1 \cdot \epsilon_2)
   \left[ 4 t (\epsilon_3\cdot k_1) (\epsilon_4\cdot k_2)
         +4 u (\epsilon_3\cdot k_2) (\epsilon_4\cdot k_1) \right]
\nonumber \\
& & \qquad
+(\epsilon_3 \cdot \epsilon_4)
   \left[ 4 t (\epsilon_1\cdot k_3) (\epsilon_2\cdot k_4)
         +4 u (\epsilon_1\cdot k_4) (\epsilon_2\cdot k_3) \right]
\nonumber \\
& & \qquad
+(\epsilon_1 \cdot \epsilon_3)
   \left[ 4 s (\epsilon_2\cdot k_3) (\epsilon_4\cdot k_1)
         +4 t (\epsilon_2\cdot k_1) (\epsilon_4\cdot k_3) \right]
\nonumber \\
& & \qquad
+(\epsilon_2 \cdot \epsilon_4)
   \left[ 4 s (\epsilon_1\cdot k_4) (\epsilon_3\cdot k_2)
         +4 t (\epsilon_1\cdot k_2) (\epsilon_3\cdot k_4) \right]
\nonumber \\
& & \qquad
+(\epsilon_1 \cdot \epsilon_4)
   \left[ 4 s (\epsilon_2\cdot k_4) (\epsilon_3\cdot k_1)
         +4 u (\epsilon_2\cdot k_1) (\epsilon_3\cdot k_4) \right]
\nonumber \\
& & \qquad
+(\epsilon_2 \cdot \epsilon_3)
   \left[ 4 s (\epsilon_1\cdot k_3) (\epsilon_4\cdot k_2)
         +4 u (\epsilon_1\cdot k_2) (\epsilon_4\cdot k_3) \right]
\Bigr\} \,.
\label{fullamp}
\ea

In the limit of low momenta (\ie\ $s,t,u$ small) the above amplitude
becomes
\be
A_4\sim   - {\kappa^{-2/3}\over 4 (4\pi)^{10/3}}
\Tr (T_1 T_2) \Tr(T_3 T_4) \frac{1}{ s d } t_8 F^4 \,.
\ee
In order to compare this to the heterotic string we need to rewrite
\be
{\kappa^{-2/3} \over d} = {\kappa^{-2/3} \over \pi R_{11} l_p}
= {(2\pi l_p)^{-3}  \over \pi R_{11} l_p}
= {2 \over (2\pi l_p)^4 R_{11}}
\ee
in string units, using the standard relation between $l_p$ and the
heterotic string length scale, $l^H_s$,
\be
l_p^2 = R_{11} \left(l^H_s\right)^2\,.
\ee
We also need  the relation between $R_{11}$ and the coupling constant,
$g_E=e^{\phi^E}$, of the ten-dimensional heterotic $E_8\times E_8$
(HE) string theory \cite{HW1},
\be
R_{11} = g_E^{2/3} = e^{{2\phi^E\over 3}} \,.
\ee
The amplitude in heterotic string units is then given by
\be
A_4^{HE} \sim  - (2\pi l^H_s)^{-4} e^{-2\phi^E} \Tr (T_1 T_2) 
\Tr(T_3 T_4) \frac{1}{s} t_8 F^4,
\ee
which agrees with the tree-level heterotic result in the low-momentum
limit \cite{ghmr,gsw}. 

This agreement should not seem surprising in view of the fact that the
low-momentum tree-level heterotic string amplitude is equal to the
tree-level amplitude of ten-dimensional $N=1$ supergravity coupled to
$E_{8\,L} \times E_{8\,R}$ $N=1$ supersymmetric Yang--Mills theory.
In the small-$R_{11}$ limit (the heterotic string weak coupling limit)
the two Ho\v{r}ava--Witten boundaries coincide and our tree-level
field theory calculation becomes ten-dimensional. The slightly subtle
point is that in the Ho\v{r}ava--Witten case the full tree amplitude
is given by the sum of `pole diagrams' which only involve three-point
vertices while the ten-dimensional supergravity calculation also
involves the contribution of a `contact' interaction of the form
$\omega_{MNP}\omega^{MNP}$ (the four gauge particle contact term does
not contribute to the particular group theory factor that is of
interest to us). In fact, it is easy to see that this contact term is
reproduced precisely by the second contribution $A_4^{(C)\, 2}$ in
appendix~B. 

\section{Compactification on $S^1$}
\setcounter{equation}{0}

Upon compactification on a circle of radius $R_9l_p$ in the $x^9$
direction the gauge group may be broken by the introduction of Wilson
lines. We will here consider the case in which the unbroken group
is $SO(16)_L \times SO(16)_R$. The two nine dimensional heterotic
strings with this gauge group are T-dual to each other. This system
has a description in Type IA language in which the Ho\v{r}ava--Witten
boundaries are represented by two orientifold eight-planes separated
by $\pi R_{11}l_p$ with eight $D8$-branes and their images placed
coincident with each of them.

The adjoint representation of $E_8$ decomposes into the adjoint and
one of the two chiral spinor representations of the $SO(16)$ subgroup,
${\bf 248}= {\bf 120}+{\bf 128}$. The breaking of the $E_8$ symmetry
is accompanied by the generation of a mass for the chiral spinor
${\bf 128}$ which is a state with Kaluza--Klein charge
$l =\pm 1/2, \pm 3/2, \dots$. The chirality of these states is
independent of the signs of the charges. In eleven-dimensional units,
the mass of the spinor states is given by
\be
M_S = {|l| \over R_9 l_p}\,.
\label{spinmass}
\ee
We will mainly consider external states that are of lowest mass, and
therefore have $l=\pm 1/2$. We shall also choose to satisfy
$M_S << M_{planck}$; in this case we may define a `low energy'
limiting effective field theory in which the spinor states survive but
Planck-scale or string-scale excitations can be ignored. In the Type
IA description a spinor state corresponds to a  single D-particle or
anti D-particle stuck on an orientifold planes. In terms of the
eleven-dimensional moduli, the Type IA string coupling constant
$g_{IA}$ is given as \cite{HW1},
\be
g_{IA} = R_9^{{3\over 2}} \,,
\ee
and the Type I string length, $l_s^I$, is given in terms of the Planck
length by
\be
{l_s^I \over l_p} = g_{IA}^{-{1\over 3}} = R_9^{-1/2}\,.
\ee
Thus in terms of the Type IA theory, the mass $M_S$ for $l=\pm 1/2$
becomes
\be\label{MS}
M_S = {1 \over 2 R_9 l_p} = {1\over l_s^I} {1\over 2 g_{IA}} \,,
\ee
which is indeed the mass of a stuck D-particle ({\it i.e.}, of a
$D_S$).

\subsection{The different kinematical regimes}

Since the $SO(16)$ spinor states carry Kaluza--Klein charge
$l = \pm 1/2$ there are three distinct classes of four-point functions
in the nine-dimensional theory. These are characterized by which of
the incoming and outgoing particles are $D_S$ and which are
$\bar D_S$. We are considering  the scattering process as a T-channel
process in which $K_1$ and $K_4$ are the nine-dimensional momenta of
the incoming particles that scatter into outgoing particles with
momenta $-K_2$ and $-K_3$. These processes are the
following:
\hfill\break\noindent
(a) $D_S + D_S \rightarrow D_S+ D_S$, which has total $S$-channel
Kaluza--Klein charge $l=0$. At low velocity this becomes a BPS
configuration that preserves supersymmetry.
\hfill\break\noindent
(b) $D_S + \bar D_S \rightarrow D_S + \bar D_S$ also has $l=0$
but is far from BPS at low velocity.
\hfill\break\noindent
(c) $D_S + \bar D_S \rightarrow \bar D_S + D_S$ has $l=1$. Although
this is far from BPS at low T-channel velocity  the S-channel process
that is related by crossing symmetry is BPS at low S-channel
velocity. In this process  a bulk D0-brane  with mass $2M_S$ is
exchanged.

The processes obtained by interchanging all $D_S$'s with $\bar D_S$'s
are trivially related to these three processes and need not be
considered separately.

We will describe the kinematics in the nine-dimensional centre of mass
frame with the outgoing states scattering through an angle $\theta$
relative to the incoming states. Writing the  ten-dimensional momentum
of the $r$'th particle as $k_r = (K_r, q_9)$, where $q_9$ is the
Kaluza--Klein momentum in the circular $x^9$ direction, gives the
explicit expressions
\be
k_1 = (E,{\bf 0},0, p ,M_S)\equiv (K_1, M_S)\,,
\label{k1def}
\ee
\be
k_2 = \left(-E',{\bf 0},-p\sin\theta,-p\cos\theta,-(-1)^l M_S\right)
\equiv (K_2, -(-1)^l M_S)\,,
\label{k2def}
\ee
\be
k_3 =\left(- E',{\bf 0},p\sin\theta,p\cos\theta, -(-1)^{l+r}M_S\right)
\equiv (K_3, -(-1)^{l+r}M_S)\,,
\label{k3def}
\ee
\be
k_4 = \left(E,{\bf 0},0,-p,(-1)^r M_S\right) \equiv (K_4,(-1)^rM_S)\,,
\label{k4def}
\ee
where ${\bf 0}$ represents the zero momentum in the six dimensions
transverse to the scattering plane, $p$ is defined by
\be
p = {M_S v \over (1-v^2)^{1/2}}\,,
\label{pdefs}
\ee
and
\be
E^2 = {M_S^2 \over (1-v^2)} = E^{\prime 2}\,.
\label{eeprdef}
\ee
The distinct choices $r =0,1$ and $0 \le l\le r$ allow for the three
processes defined above.

The nine-dimensional Mandelstam variables, $S = -(K_1+K_2)^2$,
$T = -(K_1+K_4)^2$ and $U=-(K_1+K_3)^2$ are given by
\be
S = - {2 M_S^2 v^2 \over (1-v^2)} (1-\cos\theta) \,, \qquad
 T= 4 M_S^2 + {4 M_S^2 v^2 \over (1-v^2)} \,, \qquad
U = - {2 M_S^2 v^2 \over (1-v^2)} (1+\cos\theta) \,.
\ee
These satisfy the mass-shell condition $S+T+U=4 M_S^2$.

The corresponding expressions for the ten-dimensional Mandelstam
invariants, $s,t,u$, depend importantly on which of the three kinds of
process is being considered. Given these identifications, the
nine-dimensional amplitude can be obtained  directly from the
ten-dimensional expression (\ref{totall}).

\vskip0.3cm

{\it {\bf (a)   $l=0$, $r=0$: \ \  $D_S + D_S \to D_S + D_S$}}
\vskip0.2cm

In this case the ten-dimensional kinematic invariants are
\be
s=S  \,, \qquad
t = T - 4 M_S^2 = {4 M_S^2 v^2 \over (1-v^2)} \,, \qquad
u = U\,.
\label{BPS}
\ee
This is the  process that is near-BPS at low energy. Substituting
(\ref{BPS}) into (\ref{totall}) gives an amplitude  for $D_S$
scattering proportional to
\be
{2 M_S^2 v^2 \over (1-v^2)}
\Biggl( 2 {(1+\cos(\theta)) \over (1-\cos(\theta))}
(\epsilon_1\cdot\epsilon_2) (\epsilon_3\cdot\epsilon_4)
+ 2 (\epsilon_1\cdot\epsilon_3) (\epsilon_2\cdot\epsilon_4)
- (1+\cos(\theta)) (\epsilon_1\cdot\epsilon_4)
             (\epsilon_2\cdot\epsilon_3) \Biggr) \,,
\label{BPS2}
\ee
where we have chosen polarisation vectors that satisfy
$k_r\cdot\epsilon_s=0$ for all $r,s=1,\dots,4$. This expression
vanishes in the limit of zero velocity, $v\rightarrow 0$.

In this case the amplitude is closely related to the familiar Type IIA
D-particle scattering amplitude studied in \cite{bachas,dkps}. In the
string theory description the amplitude is defined in the impact
parameter representation, where the back reaction of the interaction
on the particle trajectories is ignored. This is valid if the
D-particles are assumed to move in straight lines with slow relative
velocity $2v$ and are separated by a transverse distance $b$. In the
standard Type IIA theory the metric on the moduli space of two $D0$'s
is flat and the leading velocity-dependent term in the amplitude is
of order $v^4$.

We are here concerned with the situation in which the two D-particles
are constrained to lie on different fixed orientifold planes so the
distance of separation is at least $\pi R_{11}l_p$. In order to relate
the Feynman diagram we considered earlier to the standard D-particle
results of \cite{bachas} the amplitude must be expressed in terms of
$b$ and $v$ instead of the Mandelstam variables $S$ and $T$. This
amounts to replacing $T$ by its expression in terms of $v$ using
(\ref{BPS}) and Fourier transforming with respect to the momentum
transfer in the seven directions transverse to the particle
trajectories. In the special frame defined by
(\ref{k1def})-(\ref{k4def}) this momentum transfer has been rotated
into the vector ${\bf p} \sim ({\bf 0},p\sin\theta)$ and
$S \sim - |{\bf p}|^2$. Thus, the impact parameter representation is
defined by
\be
\tilde A_4(v, b) \sim const.
\int dS\, (-S)^{5/2}\, A_4(S,T)\, e^{i\sqrt{-S} \, b}\,,
\label{impdef}
\ee
where the overall constant includes the volume of $S^6$ to account for
the integration over angular components of the momentum transfer. The
transverse separation of the two $D_S$ particles is $|\pi R_{11}l_p+ib|$
and for $R_{11}\ne 0$ the integral in (\ref{impdef}) is dominated by the
region in which $0<< \sqrt{-S} R_{11}l_p << M_Sv$, where the factor of
$1/\sinh(\pi\sqrt{-S}R_{11}l_p)$ in $A_4$ can be approximated by
$2e^{-\pi\sqrt{-S}R_{11}l_p}$. Ignoring the deflection of the particles
amounts to picking out the leading term in the $1/|\pi R_{11}l_p+ib|$
expansion of the amplitude and it is easy to see that this leading
behaviour has a coefficient proportional to $v^4$ which comes from the
factor of $tu$ associated with the
$\epsilon_1 \cdot \epsilon_2 \, \epsilon_3 \cdot \epsilon_4$ term.

This behaviour is in agreement with the string calculation that can be
performed as in \cite{bachas}. In this case the relevant amplitudes
consist of the sum of three kinds of world-sheets. The first diagram
is the cylinder that describes the overlap between the two D-particle
boundary states representing the two $D_S$'s. The second kind of
diagram consists of the M\"obius strips that describes the overlap
involving one $D0$ boundary state and the crosscap associated with
the mirror image of that particle in the other orientifold plane.
Finally, there are the cylinder diagrams that describe the overlap of
either D-particle with any of the  sixteen $D8$-brane boundary states.
It is easy to argue that only the  $D0$-$D0$ contribution can depend
on $v$ since the velocity is constrained to be parallel to the
orientifold plane and $D8$. The $D0$-$D0$ contribution is therefore
identical to the bulk term calculated in \cite{bachas} and behaves as
$v^4$ as expected.\footnote{The leading contribution in the case of a
{\em bulk} $D0$ moving transverse to the orientifold planes is
proportional to $v^2$. This reflects the reduced amount of
supersymmetry in the system.}

\vskip0.3cm

{\it{\bf (b) $l=0$, $r=1$: \ \  $D_S +\bar D_S \to D_S +\bar D_S$}}

\vskip0.2cm

In this case the ten-dimensional kinematic invariants are
given by
\be
s=S \, , \qquad
t = T  \,, \qquad
u = U - 4 M_S^2 = - 4 M_S^2
             - {2 M_S^2 v^2 \over (1-v^2)} (1+\cos\theta) \,.
\label{notu}
\ee
The non-BPS amplitude that results by substituting these expressions
into (\ref{totall}) behaves to leading order in $v$ as
\be
{8 M_S^2 (1-v^2) \over v^2 (1-\cos(\theta))}
(\epsilon_1\cdot\epsilon_2) (\epsilon_3\cdot\epsilon_4) \,,
\ee
which diverges as $v\rightarrow 0$. This divergence is presumably
related to the instability of the system to decay into a non-BPS
D-string \cite{BGH,DS} which will not be discussed further here.

\vskip0.3cm

{\it{\bf  (c)  $l= 1$, $r=1$: \ \  $D_S + \bar D_S \to \bar D_S + D_S$ }}

\vskip0.2cm

The ten-dimensional kinematic invariants are
\be
s = S - 4 M_S^2 = - 4 M_S^2
                  - {2 M_S^2 v^2 \over (1-v^2)} (1-\cos\theta)\,, \qquad
t=T\,,\qquad u=U\,.
\label{tudefs}
\ee
This is the case in which there is $D0$ exchange and the behaviour of
the amplitude is qualitatively different from the cases with $l=0$
since $s\rightarrow -4M_S^2$ is large in the low-velocity limit,
$v\rightarrow 0$. The denominator of (\ref{totall}) can then be
expanded as
\be\label{eq18}
{1 \over \sinh(\pi\sqrt{-s}R_{11}l_p)} = 2 e^{-2 \pi M_S R_{11}l_p}
\sum_{r=0}^\infty e^{-4 \pi r M_S R_{11}l_p} \,.
\ee
Using the expression for $M_S$ from (\ref{MS}), (\ref{eq18}) then
becomes
\be\label{result}
2 \sum_{r=0}^{\infty} e^{-(2r+1)\pi {R_{11}} \over R_{9}}
= 2 \sum_{r=0}^{\infty} e^{-(2r+1)\pi {R^{IA}\over g_{IA}}}
= 2 \sum_{r=0}^{\infty} e^{- {(2r+1) \pi  \over g_I}} \,,
\ee
where we have used the standard relation between the M-theory
variables and the ten-dimen\-sio\-nal Type IA coupling constant,
$g_{IA} = e^{\phi^{IA}}$, and radius, $R_{IA}$,
\be
R_{IA} = R_{11}R_9^{1/2} , \qquad g_{IA} = R_9^{3/2},
\label{raddef}
\ee
together with the familiar T-duality relation between Type IA and
Type I,
\be
g_I = {g_{IA} \over R_{IA}}= {R_9 \over R_{11}} \,, \qquad
R_I = {1 \over R_{IA}} = {1\over R_{11} R_9^{1/2}} \,.
\ee
In order to rewrite the total amplitude (\ref{totall}) in terms of
Type IA (or Type I) variables, we also have to re-express the
dimensionful quantities in terms of the string scale. As before, the
amplitude of the original external states is proportional to
$\kappa^{-2/3}$ which is now equal
to
\be
\kappa^{-2/3} = {1 \over (2\pi l_p)^3}
= {1 \over (2\pi l_s^I)^3} e^{-\phi^{IA}} \,.
\ee
Together with $\sqrt{-s}= 2 M_S = (g_{IA} l_s^I)^{-1}$ the amplitude
thus becomes
\be
A_4^I = {1 \over (2\pi l_s^I)^2} {1 \over (4 \pi)^{13/3}}
\Tr (T_1 T_2) \Tr(T_3 T_4) t_8 F^4
\sum_{r=0}^{\infty} e^{- {(2r+1) \pi  \over g_I}} \,.
\label{finamp}
\ee

This expansion as a series of exponentially suppressed terms is
relevant at weak Type I coupling, $g_I= R_{9}/R_{11} << 1$, where the
leading term dominates. This term can be identified as the effect of a
non-BPS D-instanton with action $\pi/g_I$.\footnote{As we shall
demonstrate in the next section, this process also exhibits the key
characteristic of the D-instanton of Type I: it breaks $O(32)$
to $SO(32)$.} According to \cite{WittenK,Frau} the Type I D-instanton
can be thought of as the sum of the Type IIB D-instanton and the
anti-D-instanton (without any prefactors). The orientation-reversing
operator $\Omega$ maps the D-instanton boundary state to the
anti-D-instanton boundary state, and therefore only the  sum of the
two boundary states is invariant. Under T-duality, the D-instanton
becomes a $D0$-brane whose world-line stretches across the  interval
between the orientifold planes of the nine-dimensional Type IA theory,
and the anti-D-instanton becomes the configuration in which the
world-line has the opposite orientation. Again this can be represented
by two boundary states that are mapped into one another under the
T-dual of $\Omega$, $\Omega\I_{11}$. The D-particle world-line
transfers  $D0$ Ramond--Ramond charge from one orientifold plane to
the other, while the image under $\Omega \I_{11}$ transfers the
opposite $D0$ Ramond-Ramond charge. In the process
$D0+ \bar D0\to\bar D0+D0$ only of these two processes contributes;
the conjugate process $\bar D0+D0 \to  D0 + \bar  D0$ would pick out
the other contribution.

In our conventions, the mass of a bulk D-particle in Type IA is
$1/g_{IA}$, and comparison with Type IIA gives
$g_{IA}=g_{IIA}$. On the other hand, under T-duality, the stuck
D-particle of Type IA (with mass $1/(2 g_{IA})$) becomes a D1-brane of
Type I that wraps the circle $R_I$ once.  This has a  mass
$R_I/(2g_I)$ and therefore $2 g_I = g_{IIB}$. Using this,
the action of the type I  D-instanton, that is the modulus of the
exponent of the $r=0$ term in (\ref{finamp}), has the value
\be
{\pi \over g_I} = {2\pi \over g_{IIB}} \,,
\ee
in Type IIB units. This is precisely the action of a single
D-instanton of Type IIB \cite{greengut}. The agreement is a
consequence of the BPS nature of the exchanged type IIA D-particle. 

The terms with $r>0$ in (\ref{finamp}) describe processes where the
D-particle is emitted on one boundary and absorbed at the other, but
where it winds an integer number of times around the compact
circle. These terms appear as multi-instanton contributions with  odd
D-instanton number. However, all of these multi-instanton terms carry
the same topological $\Zop_2$ charge, and the contributions with $r>0$
are thus presumably `unstable' and therefore only the leading
contribution with $r=0$ should be taken seriously in the perturbative
Type I limit. However, the complete series is crucial in determining
the amplitude in the limit $g_I \to \infty$, which is the weak
coupling limit of the $SO(32)$ heterotic string theory (the HO
theory) compactified on a circle with appropriate Wilson lines to
break the symmetry to $SO(16) \times SO(16)$. The heterotic coupling
is given by $g_{HO}= 1/g_I$ so that the amplitude (\ref{finamp})
has an infinite power series expansion in heterotic string
perturbation theory. After taking into account the rescaling of the
string scale
\be
l_s^I = l_s^H \left( {R_{11} \over R_{9}} \right)^{1/2}
      = l_s^H g_{HO}^{1/2}\,,
\label{scaledef}
\ee
needed to pass from the Type I to the heterotic $SO(32)$ theory, the
amplitude has an expansion in powers of $g_{HO}^2$,
\be
A_4^{HO} \sim {1 \over g_{HO}^2 (2\pi l_s^H)^2}
\Tr (T_1 T_2) \Tr(T_3 T_4)
t_8 F^4 (1 +O(g^2_{HO}))\,,
\label{finag}
\ee
where we have used
\be
\sum_{r=0}^{\infty} e^{-(2r+1) \pi \over g_I}
= {1 \over 2 \sinh(\pi g_{HO})} = {1\over 2\pi g_{HO}}
(1 +O(g^2_{HO}))\,.
\ee
The first term in (\ref{finag}) is a tree-level term that can be
determined directly from the tree-level heterotic string compactified
on a circle with the appropriate Wilson lines to give the unbroken
$SO(16) \times SO(16)$.  The $O(g^2_{HO})$ corrections appear to 
correspond to  loop corrections to the amplitude, but these
might also get contributions from M-theory loops that have not been
considered in this paper. 

The M-theory tree-level calculation can be trusted whenever both
$R_{11}$ and $R_{9}$ are large. In particular, this implies that the
ten-dimensional Type IA coupling is large, and that the radius of the
Type I theory is small. On the other hand, the ten-dimensional Type I
coupling only depends on the ratio of $R_{9}$ and $R_{11}$, and
therefore can be small. However this does not imply that one can trust
the Type I perturbation. In fact, $R_I = g_I/g_{IA} << g_I$ and
therefore the condensation of closed-string winding states cannot be
ignored. This can also be seen from the fact that both the radius and
the ten-dimensional coupling constant of the dual Type IA theory are
large.

If we had been working in a regime in which perturbative string theory
could be trusted there would have been a paradox. We would have been
led to believe that the Type I D-instanton is only a `stable'
solution (\ie, a solution with no tachyonic modes) provided the radius
$R_{I}$ is sufficiently large. Within the perturbative approximation,
when $R_{I}<\sqrt{2}$ the open string beginning and ending on the
D-instanton develops a tachyonic mode and the D-instanton should
`decay' into a non-BPS $D0$-$\bar D0$ pair whose world-lines stretch
along the circle.\footnote{The K-theory class that corresponds to the
D-instanton is however non-trivial for all values of the radii. The
question of stability is the question of whether the representative of
the non-trivial K-theory class that has least action is a pointlike
instanton or an extended (one-dimensional) object.} Furthermore, these
$D0$ and $\bar D0$ branes carry a relative $\Zop_2$ Wilson line. Under
T-duality, the two $\Zop_2$ $D0$ branes become two $\Zop_2$
D-instantons of Type IA, and the relative Wilson line means that they
are located at opposite orientifold planes. From the point of view of
Type IA, the above superposition of $D0$ world-lines would therefore
be unstable to decay into two non-BPS Type IA D-instantons provided 
that $R_{IA}>1/\sqrt{2}$. It is easy to see that such a configuration
is not at all similar to the exchange of a D-particle that is
contained in the M-theory amplitude.

However, the region of validity of our M-theory argument is one of
strong coupling  in the Type IA theory, in which the preceding
stability arguments are not valid. The fact that the D-instanton
action extracted from the spinor scattering process corresponds to
that expected for large $R_I$ in the Type I theory suggests that the 
non-BPS Type I D-instanton remains stable in the non-perturbative
region $R_I<< g_I$. The reason why the M-theory calculation reproduces
the correct value for the instanton action is related to the fact that
for the process we are considering, only one term in the superposition
of the D-instanton and the anti-D-instanton contributes. The
corresponding D-instanton action is therefore protected by
supersymmetry since the bulk D0-brane that is exchanged is a BPS state
of Type IIA (and the anti-D0-brane does not play a role). We thus
expect that the exponent in the amplitude can be trusted beyond the
original regime of validity. Although the limit which gives
ten-dimensional weakly coupled Type I theory reproduces the correct
normalisation of the instanton action it will not give the correct
value for the coefficient of the exponential.

\subsection{The Type I D-instanton and the breaking of $O(32)$}

{}From its description in terms of K-theory, it is clear that the
D-instanton is associated with the fact that the gauge group of Type I
string theory is $SO(32)$ rather than the $O(32)$ that might have been
expected on the basis of perturbation theory
\cite{WittenK}.\footnote{The actual gauge group is obviously 
$Spin(32)/\Zop_2$. The issue of the correct spin cover will be
discussed later.} Here we will demonstrate this by considering the
dual Type IA theory in the standard $SO(16)\times SO(16)$ vacuum.
Consideration of the perturbative approximation to the low energy
effective field theory would suggest that the symmetry group could be
$O(16)\times O(16)$. However, we will see that the instantonic
contribution to the scattering of spinor states is not invariant under
the disconnected component of either of the two $O(16)$ groups.

We shall concentrate on the $O(16)$ group that is associated to the
left orientifold plane since the argument for the other gauge group is
identical. In the vacuum we are considering, eight $D8$-branes (plus
eight mirror $D8$-branes) are located at each of the two orientifold
planes. As we have seen, the instantonic contribution to the
scattering amplitude comes from the diagram where a $D0$ (or
$\bar{D0}$) is emitted from the left orientifold plane and absorbed at
the right orientifold plane. In this process, the $D0$ (and its mirror
partner) has to cross the eight $D8$-branes (and their mirror
partners) that are localised at the fixed plane. However, whenever a
$D0$ crosses a $D8$-brane, a fundamental string is created that
stretches between the $D0$ and the $D8$-brane \cite{08}. This is an
important effect as we shall show momentarily.

In order to obtain a clear picture, it is useful to consider the
configuration where the eight $D8$-branes have been moved to 
$x^{11}=\epsilon$, away from the orientifold fixed plane at $x^{11}=0$
(with corresponding displacements of the mirror eight-branes to
$x^{11}=-\epsilon$). In this configuration, the `cosmological
constant' $m$ takes the value $m=-8$ for $0< x^{11} < \epsilon$ and
$m=0$ for $|x^{11}| > \epsilon$. As was shown by \cite{polstrom}, in a
background characterised by $m$, each $D0$ has to carry $|m|$ strings,
where the sign of $m$ determines whether the string begins or ends on 
the $D0$. The $D0$ (and its mirror) that are emitted from the
orientifold plane each carry eight strings in the regime 
$|x^{11}| < \epsilon$. These strings can be thought of as stretching
between the bulk $D0$ (that has been emitted from the orientifold
plane) and the remaining fractional $D0$ that is still stuck. As the
bulk $D0$ (and its mirror) crosses the eight $D8$-branes (and their
mirrors), additional strings are created that join it to the
$D8$-branes. These can recombine with the original strings to give
eight strings that stretch between the stuck $D0$ and the eight
$D8$-branes, as well as eight strings that stretch between the stuck
$D0$ and the eight mirror $D8$-branes.  In the final configuration the
bulk $D0$ does not have any strings attached to it, which is
consistent with the fact that $m=0$ in the interval between the two
fixed planes. 

The sixteen different strings that stretch between the stuck $D0$
and the eight plus eight $D8$-branes can be identified with the
generators $\gamma_i$, $i=1,\ldots,16$ of the Clifford algebra
\be
\{ \gamma_i,\gamma_j \} = 2 \delta_{ij}
\ee
that give rise to the Lie algebra elements of $so(16)$ by
\be
\sigma_{ij} = {1\over 4} [\gamma_i,\gamma_j] \,.
\ee
Indeed, the elements of the gauge group correspond to 8-8 strings, and
are therefore bilinear in the 0-8 strings that are described by the
$\gamma_i$. The product of all sixteen 0-8 strings (that arises
naturally in the above configuration) is then the {\em chirality
operator} of $O(16)$,
\be
\Gamma = \prod_{i=1}^{16} \gamma_i \,.
\ee
Thus it follows that the vertex that describes the emission of a
single $D0$ (or $\bar D0$) is actually given by
\be
\Tr \left( S_1 S_2 \Gamma \right) \,,
\ee
where $S_i$ describes the state in the spinor representation for the
two fractional $D0$'s. More generally, if $m$ bulk $D0$'s are
emitted, the vertex will be
\be
\Tr \left( S_1 S_2 \Gamma^m \right) \,,
\ee
since each bulk $D0$ gives rise to one chirality operator.

The chirality operator $\Gamma$ is invariant under conjugation by any
element of $SO(16)$, but it changes sign under conjugation by an
element $g_0$ in the disconnected component of $O(16)$,
$g_0 \Gamma g_0^{-1} = - \Gamma$. This implies that the vertex with
odd $m$ only respects the $SO(16)$ subgroup of $O(16)$. Since the
contributions with odd $m$ are precisely those that appear in the
instantonic contribution to the amplitude, this exponentially
suppressed contribution is associated with  the breaking of $O(16)$ to
$SO(16)$. In terms of Type I, $SO(16) \times SO(16)$ is a subgroup of
$SO(32)$. The large $O(32)$ transformations that are not in $SO(32)$
can be taken to lie in $O(16)\times SO(16)$; these transformations are
indeed not respected by the D-instanton, as was stressed in
\cite{WittenK}.

The actual gauge group of the ten-dimensional Type I theory is
$Spin(32)/\Zop_2$. Upon compactification to nine dimensions with the
above Wilson line, this is broken to
$S(Pin(16)\times Pin(16))/\Zop_2$, where the $S$ outside the bracket
indicates that either both the elements of $Pin(16)\times Pin(16)$
are in the disconnected component of $Pin(16)$ or neither of them
are. As we have just explained, the effect of the D-instanton of Type
I is to impose this constraint. The actual gauge group is somewhat
smaller since the nine-dimensional Type IA theory also has a
D-instanton that is T-dual to the wrapped D0-brane of Type I. This
instanton is stuck on one of the two fixed planes, and it is
responsible for breaking each $Pin(16)$ separately to
$Spin(16)$. (This follows by essentially the same arguments that were
used by Witten for the case of the ten-dimensional Type I
D-instanton.) Thus the actual gauge group is 
\be
\left( Spin(16) \times Spin(16) \right) / \Zop_2 \,,
\ee
where the element in the centre by which the group is divided is the
element that does not allow states transforming in the vector
representation of one $Spin(16)$ and the scalar representation of the
other. It is not difficult to see that all allowed (and no other)
conjugacy classes of representations of this gauge group are actually
present in both nine dimensional heterotic theories. (This discussion
is relevant for some issues raised in \cite{McI}.)

\section{Discussion}

In this paper we have identified the D-instanton contribution to a
certain amplitude involving spinor states of Type I, using the
relation of nine-dimensional Type I to M-theory. This relation
involves Type IA, and it is therefore natural to ask whether
the $\Zop_2$ non-BPS D-instanton of Type IA (that is T-dual to the
D0-brane of Type I with its world-line wrapped along $x^9$) can also
be understood in terms of M-theory (compare also \cite{Mich}). The
wrapped D0-brane of Type I can be obtained from a D1-brane
anti-D1-brane pair that are wrapped around $x^9$ and $x^8$ say, and
that have a relative Wilson line along $x^8$. Under T-duality of the
$x^9$-circle, we therefore obtain a D0-brane and an anti-D0-brane,
both localised on the {\em same} orientifold plane and wrapped around
$x^8$ with a relative Wilson line. One should therefore expect that
the Type IA D-instanton contributes to the scattering diagram 
$D_S+ \bar D_S \to \bar D_S + D_S$ where now all four D-particles are
localised on the same boundary.  An analysis of this process can be
made that is analogous to that of the main part of the paper, with the
important  difference that the propagator (\ref{schw1}) does not
involve a $(-1)^m$, and therefore becomes
\be
\widehat{\cal G} (p^M; 0, 0)  =  {1 \over \pi R_{11}l_p}
\sum_{m=-\infty}^\infty {1 \over p^2 + {m^2\over R_{11}^2l_p^2}}
= {1 \over \sqrt{p^2}} \sum_{n\in\Zop}
e^{-2 |n| \pi \sqrt{p^2} R_{11}l_p} \,.
\ee
The term with $n=0$ should correspond to a part of the contribution
of the Type IA $\Zop_2$ D-instanton. The terms with $n\ne 0$ come
from multiple windings of the $D0$ world-line and, as in the earlier
case, carry the same $\Zop_2$ topological charge as the $n=0$ term;
the latter is therefore the dominant contribution. Thus the 
tree amplitude describing the exchange of the graviton and the
three-form in Type IA variables is approximated by 
\be
A_4 = {1\over 2 (2\pi l_s^I)^2} {1\over (4\pi)^{13/3}}
t_8 F^4 \left[ \Tr (T_1 T_2) \Tr(T_3 T_4) + \cdots \right] \,,
\label{oneaa}
\ee
where we have only exhibited explicitly the term whose group structure
is the same as in the main part of the paper. By essentially the same
arguments as above, the coupling of the spinors to the bulk graviton
and three-form should again involve a chirality operator and therefore
break $O(16)$ to $SO(16)$. However, the Type IA instanton should
contribute a term of the form $e^{-c/g_{IA}}$ to the amplitude (for
some constant, $c$).  Since the M-theory calculation is justified only
in the region of large large $R_9, R_{11}$, where  $g_{IA}>>1$, the
expression (\ref{oneaa}) can only represent the first term in an
expansion of the exponent in powers of $1/g_{IA}$. The higher-order
terms will depend on the undetermined loop diagrams of M-theory.

\section*{Acknowledgements}

We acknowledge partial support from the PPARC SPG programme, ``String
Theory and Realistic Field Theory'', PPA/G/S/1998/00613. TD is
supported by a Scholarship of Trinity College, Cambridge, and MRG is 
supported by a Royal Society University Research Fellowship. MRG and
MBG are grateful to the California Institute of Technology for
hospitality during the final stages of this work. We also thank 
Oren Bergmann and Michael Dine for useful conversations.

\section*{Appendix}

\appendix

\section{The Ho\v{r}ava--Witten action}

\setcounter{equation}{0}

The spinor indices are written as $\alpha, \beta, \gamma$. The
supergravity  multiplet has the graviton $g$, the gravitino
$\psi_{\mu\alpha}$, and a  three-form $C$ with the field strength $G$
(in component form
$G_{\mu\nu\rho\sigma}=4!\partial_{[\mu}C_{\nu\rho\sigma]}$). The
spinors are Majorana and $\overline{\psi}_\alpha$ is defined by
$\overline{\psi}_\alpha = C_{\alpha\beta}\psi^\beta$ where
$C_{\alpha\beta}$ is the charge conjugation matrix. The $32 \times 32$
real Dirac matrices $\Gamma_\mu$ satisfy the Clifford algebra
$\{\Gamma_\mu,\Gamma_\nu\} = 2g_{\mu\nu}$, and we define
$\Gamma^{\mu_1\mu_2 \ldots \mu_n} \equiv
{1\over n!}\Gamma^{\mu_1}\Gamma^{\mu_2}\ldots \Gamma^{\mu_n} \pm$
permutations. With these conventions the $D=11$, $N=1$ bulk
supergravity action is given by \cite{cjs},
\ba
S_{bulk} &=& \frac{1}{\kappa^2}\int_{R^{1,9} \times S^1} d^{11}x
\sqrt{-g} \Bigg\{-\frac{R}{2}-{1 \over 2}\overline{\psi}_\mu
\Gamma^{\mu\nu\rho} D_\nu \psi_\rho
- \frac{1}{48} G_{\mu\nu\rho\sigma}G^{\mu\nu\rho\sigma}  \\
& &  - \frac{\sqrt{2}}{192} ( \overline{\psi}_\mu
\Gamma^{\mu\nu\rho\sigma\tau\lambda}
\psi_\lambda + 12\overline{\psi}^\nu
\Gamma^{\rho\sigma}\psi^\tau ) G_{\nu\rho\sigma\tau}
- \frac{\sqrt{2}}{3456}
\epsilon^{\mu_1\mu_2\ldots \mu_{11}} C_{\mu_1\mu_2\mu_3}
G_{\mu_4 \ldots \mu_7} G_{\mu_8 \ldots \mu_{11}}\Bigg\}\,. \nonumber
\label{Sbulk}
\ea
Here $\kappa^2=(2\pi l_p)^9$, where $l_p$ is the eleven-dimensional
Planck length. We are not considering higher order terms quartic in
the gravitino. The Riemann tensor is the field strength of the spin
connection $\Omega$. On the other hand the D=10 supersymmetric
Yang-Mills action on each boundary, containing the $E_8$ gauge field
$A^a$ and the gluino $\chi^a$, coupled to the bulk supergravity
fields, is given by \cite{HW2},
\ba
S_{YM} & = &  {1 \over (4\pi)^{5/3}\kappa^{4/3}} \int_{R^{1,9}} d^{10}x
    \sqrt{-g} \Bigg[-\frac{1}{4} F^a_{MN} F^{aMN}
    -\frac{1}{2} {\bar{\chi}}^a \Gamma^M D_M \chi^a \nonumber  \\
& & -\frac{1}{4} {\bar{\psi}}_M \Gamma^{NP}\Gamma^M F^a_{NP} \chi^a
     +{\bar{\chi}}^a \Gamma^{MNP} \chi^a
         \Bigg\{ \frac{\sqrt2}{48} G_{MNP11}
            +\frac{1}{32} {\bar{\psi}}_M \Gamma_{NP}\psi_{11}
            +\frac{1}{32} {\bar{\psi}}^D \Gamma_{DMNP} \psi_{11}
\nonumber \\
& & +\frac{1}{128} \big(3 {\bar{\psi}}_M \Gamma_N \psi_P
          - {\bar{\psi}}_M \Gamma_{NPQ} \psi^Q
          -\frac{1}{2}{\bar{\psi}}_Q \Gamma_{MNP}\psi^Q
          -\frac{13}{6}{\bar{\psi}}^D \Gamma_{QMNPR}\psi^R \big)
   \Bigg\} \Bigg]\,,
\label{Sym}
\ea
where $g$ is here the restriction of the eleven-dimensional metric to
ten dimensions, and the Yang-Mills field strength is given by
$F^a_{MN} =\partial_M A^a_N - \partial_N A^a_M + f^a_{bc}A^b_MA^c_N$.
The Yang-Mills coupling $\lambda$ has been expressed in terms of the
gravitational coupling $\kappa$ by the relation
$\lambda^2 = 4\pi (4\pi \kappa^2)^{2/3}$ \cite{HW2}; the prefactor is
therefore $(4\pi)^{5/3} (2\pi l_p)^6$. The expression for
$G_{MNP\,11}$ is modified in the manner described by (\ref{Ghw}).

\section{The separate tree amplitudes}

\setcounter{equation}{0}

The tree amplitude due to graviton exchange is given by
\ba
A_4^{(h)} & = &  {\kappa^{-2/3}\over (4\pi)^{10/3} d}
\Tr(T_1 T_2) \Tr(T_3 T_4)
\sum_{m\in\Zop} (-1)^m {1 \over -s + p^2_{11}} \nonumber \\
& &  \qquad \times
\left[ \epsilon_1^{N_1} k_1^{M_1} 
\epsilon_2^{Q_1} k_2^{P_1}
+ \epsilon_1^{Q_1} k_1^{P_1} \epsilon_2^{N_1} k_2^{M_1}
\right]
\nonumber \\
& & \qquad \times \Big(\half\eta_{M_1[Q_1}\eta_{P_1]N_1}\eta_{R_1S_1} +
\eta_{M_1[P_1}\eta_{Q_1]S_1}\eta_{R_1N_1}+
\eta_{M_1[R_1}\eta_{Q_1]N_1}\eta_{P_1S_1}\Big)
\nonumber \\
& & \qquad \times \Bigg(\eta^{R_1 R_2}\eta^{S_1S_2}
+ \eta^{R_1S_2} \eta^{S_1R_2}
- \frac{2}{9}\eta^{R_1S_1}\eta^{R_2S_2}\Bigg)
\nonumber \\
& & \qquad \times \Big(\half\eta_{M_2[Q_2}\eta_{P_2]N_2}\eta_{R_2S_2} +
\eta_{M_2[P_2}\eta_{Q_2]S_2}\eta_{R_2N_2}+
\eta_{M_2[R_2}\eta_{Q_2]N_2}\eta_{P_2S_2}\Big)
\nonumber \\
& & \qquad \times
\left[ \epsilon_3^{N_2} k_3^{M_2} 
       \epsilon_4^{Q_2} k_4^{P_2}
+ \epsilon_3^{Q_2} k_3^{P_2} 
  \epsilon_4^{N_2} k_4^{M_2}
\right] \,.
\label{newb}
\ea
The sum over $m$ can be performed directly using (\ref{finalprop1}),
and after a lengthy calculation the result simplifies to
\ba
A_4^{(h)} &=&  {\kappa^{-2/3}\over 4 (4\pi)^{10/3}}
\Tr(T_1 T_2) \Tr(T_3 T_4)
{1 \over \sqrt{-s}} {1 \over \sinh(\sqrt{-s} d)} \Bigl\{
\nonumber \\
& & \quad
-2ut (\epsilon_1 \cdot \epsilon_2) (\epsilon_3 \cdot \epsilon_4)
+ s^2 (\epsilon_1 \cdot \epsilon_3) (\epsilon_2 \cdot \epsilon_4)
+ s^2 (\epsilon_1 \cdot \epsilon_4) (\epsilon_2 \cdot \epsilon_3)
\nonumber \\
& & \quad + (\epsilon_1 \cdot \epsilon_2)
 \Bigl[  4t (\epsilon_3\cdot k_1) (\epsilon_4\cdot k_2)
        +4u (\epsilon_3\cdot k_2) (\epsilon_4\cdot k_1) \Bigr]
\nonumber \\
& & \quad + (\epsilon_3 \cdot \epsilon_4)
 \Bigl[  4t (\epsilon_1\cdot k_3) (\epsilon_2\cdot k_4)
        +4u (\epsilon_1\cdot k_4) (\epsilon_2\cdot k_3) \Bigr]
\nonumber \\
& & \quad + (\epsilon_2 \cdot \epsilon_3)
 \Bigl[ +2s (\epsilon_1\cdot k_4) (\epsilon_4\cdot k_3)
        +2s (\epsilon_1\cdot k_2) (\epsilon_4\cdot k_1)
        -2t (\epsilon_1\cdot k_2) (\epsilon_4\cdot k_3)\Bigr]
\nonumber \\
& & \quad + (\epsilon_1 \cdot \epsilon_3)
 \Bigl[ +2s (\epsilon_2\cdot k_4) (\epsilon_4\cdot k_3)
        +2s (\epsilon_2\cdot k_1) (\epsilon_4\cdot k_2)
    -2u (\epsilon_2\cdot k_1) (\epsilon_4\cdot k_3)\Bigr]
\nonumber \\
& & \quad + (\epsilon_2 \cdot \epsilon_4)
 \Bigl[ +2s (\epsilon_1\cdot k_3) (\epsilon_3\cdot k_4)
        +2s (\epsilon_1\cdot k_2) (\epsilon_3\cdot k_1)
        -2u (\epsilon_1\cdot k_2) (\epsilon_3\cdot k_4)\Bigr]
\nonumber \\
& & \quad + (\epsilon_1 \cdot \epsilon_4)
 \Bigl[ +2s (\epsilon_2\cdot k_3) (\epsilon_3\cdot k_4)
        +2s (\epsilon_2\cdot k_1) (\epsilon_3\cdot k_2)
        -2t (\epsilon_2\cdot k_1) (\epsilon_3\cdot k_4)\Bigr]
\nonumber \\
& & \quad
- 4 (\epsilon_1 \cdot k_2) (\epsilon_2\cdot k_1)
       (\epsilon_3 \cdot k_4) (\epsilon_4\cdot k_3)
+ 4 (\epsilon_1 \cdot k_4) (\epsilon_2\cdot k_1)
       (\epsilon_3 \cdot k_2) (\epsilon_4\cdot k_3)
\nonumber \\
& & \quad
+ 4 (\epsilon_1 \cdot k_2) (\epsilon_2\cdot k_4)
       (\epsilon_3 \cdot k_1) (\epsilon_4\cdot k_3)
+ 4 (\epsilon_1 \cdot k_3) (\epsilon_2\cdot k_1)
       (\epsilon_3 \cdot k_4) (\epsilon_4\cdot k_2)
\nonumber \\
& & \quad
+ 4 (\epsilon_1 \cdot k_2) (\epsilon_2\cdot k_3)
       (\epsilon_3 \cdot k_4) (\epsilon_4\cdot k_1) \Bigr\} \,.
\label{b2}
\ea
The  three-form exchange diagram involves a vertex of the form
$S^{AAC}$ (\ref{aacvert}) at each end of the propagator, and
thus the total amplitude is of the form
\ba
A_4^{(C)} &= & {2 \kappa^{-2/3} \over (4\pi)^{10/3} (3!)^2 d}
\Tr(T_1 T_2) \Tr(T_3 T_4)
\sum_{m\in\Zop} (-1)^m {1 \over -s + p^2_{11}} \nonumber \\
& & \left[\epsilon_1^{M_1} k_2^{N_1} \epsilon_2^{P_1}
+ \epsilon_2^{M_1} k_1^{N_1} \epsilon_1^{P_1} \right]
\nonumber \\
& &
\left(p^{[M_2} p_{[M_1} \eta_{N_1 P_1} \eta_{11]}^{N_2}
\eta^{P_2\, 11]} \right)
\left[\epsilon_3^{M_2} k_4^{N_2} \epsilon_4^{P_2}
+ \epsilon_4^{M_2} k_3^{N_2} \epsilon_3^{P_2} \right] \,,
\label{b3}
\ea
where $p=(k_1+k_2,p_{11})$, since the external gauge bosons do not
have any momentum (or polarisation vectors) in the 11th
direction. The propagator therefore has two contributions
\be
\left(3 k_{[M_2} k^{[M_1} \eta^{N_1 P_1]} \eta_{N_2 P_2]}
- p_{11}^2 \eta^{[M_1 N_1} \eta^{P_1]}_{[M_2}
\eta_{N_2 P_2]} \right) \,,
\ee
where $k=k_1+k_2=k_3+k_4$. In the first term we can again do the sum
over $p_{11}$ as before, and we obtain a contribution of the form
\ba
A_4^{(C)\, 1} &= & {\kappa^{-2/3}\over 4 (4\pi)^{10/3}}
\Tr(T_1 T_2) \Tr(T_3 T_4)
{1 \over \sqrt{-s}} {1 \over \sinh( \sqrt{-s} d)} \Bigl\{
\nonumber \\
& & \quad
+ (u-t) (\epsilon_1\cdot\epsilon_4) (\epsilon_2\cdot k_1)
     (\epsilon_3\cdot k_4)
+ (t-u) (\epsilon_1\cdot\epsilon_3) (\epsilon_2\cdot k_1)
     (\epsilon_4\cdot k_3)
\nonumber \\
& & \quad
+ (t-u) (\epsilon_2\cdot\epsilon_4) (\epsilon_1\cdot k_2)
     (\epsilon_3\cdot k_4)
+ (u-t) (\epsilon_2\cdot\epsilon_3) (\epsilon_1\cdot k_2)
     (\epsilon_4\cdot k_3)
\nonumber \\
& & \quad
+ (\epsilon_1\cdot k_3) (\epsilon_2\cdot k_1) (\epsilon_3\cdot k_4)
         (\epsilon_4\cdot k_1)
- (\epsilon_1\cdot k_3) (\epsilon_2\cdot k_1) (\epsilon_3\cdot k_4)
         (\epsilon_4\cdot k_2)
\nonumber \\
& & \quad
- (\epsilon_1\cdot k_4) (\epsilon_2\cdot k_1) (\epsilon_3\cdot k_4)
         (\epsilon_4\cdot k_1)
+ (\epsilon_1\cdot k_4) (\epsilon_2\cdot k_1) (\epsilon_3\cdot k_4)
         (\epsilon_4\cdot k_2)
\nonumber \\
& & \quad
+ (\epsilon_1\cdot k_4) (\epsilon_2\cdot k_1) (\epsilon_3\cdot k_1)
         (\epsilon_4\cdot k_3)
- (\epsilon_1\cdot k_4) (\epsilon_2\cdot k_1) (\epsilon_3\cdot k_2)
         (\epsilon_4\cdot k_3)
\nonumber \\
& & \quad
- (\epsilon_1\cdot k_3) (\epsilon_2\cdot k_1) (\epsilon_3\cdot k_1)
         (\epsilon_4\cdot k_3)
+ (\epsilon_1\cdot k_3) (\epsilon_2\cdot k_1) (\epsilon_3\cdot k_2)
         (\epsilon_4\cdot k_3)
\nonumber \\
& & \quad
+ (\epsilon_1\cdot k_2) (\epsilon_2\cdot k_3) (\epsilon_3\cdot k_4)
         (\epsilon_4\cdot k_2)
- (\epsilon_1\cdot k_2) (\epsilon_2\cdot k_3) (\epsilon_3\cdot k_4)
         (\epsilon_4\cdot k_1)
\nonumber \\
& & \quad
- (\epsilon_1\cdot k_2) (\epsilon_2\cdot k_4) (\epsilon_3\cdot k_4)
         (\epsilon_4\cdot k_2)
+ (\epsilon_1\cdot k_2) (\epsilon_2\cdot k_4) (\epsilon_3\cdot k_4)
         (\epsilon_4\cdot k_1)
\nonumber \\
& & \quad
+ (\epsilon_1\cdot k_2) (\epsilon_2\cdot k_4) (\epsilon_3\cdot k_2)
         (\epsilon_4\cdot k_3)
- (\epsilon_1\cdot k_2) (\epsilon_2\cdot k_4) (\epsilon_3\cdot k_1)
         (\epsilon_4\cdot k_3)
\nonumber \\
& & \quad
- (\epsilon_1\cdot k_2) (\epsilon_2\cdot k_3) (\epsilon_3\cdot k_2)
         (\epsilon_4\cdot k_3)
+ (\epsilon_1\cdot k_2) (\epsilon_2\cdot k_3) (\epsilon_3\cdot k_1)
         (\epsilon_4\cdot k_3) \Bigr\} \,.
\label{b5}
\ea
In the second term, the sum over $m$ is evaluated using (\ref{schw2}),
and we obtain
\ba
A_4^{(C)\, 2} &= &  {\kappa^{-2/3}\over 4 (4\pi)^{10/3}}
\Tr(T_1 T_2) \Tr(T_3 T_4) \sqrt{-s} {1 \over \sinh(\sqrt{-s} d)}
\Bigl\{
\nonumber \\
& & \quad
+ (t-u) (\epsilon_1\cdot\epsilon_3) (\epsilon_2\cdot\epsilon_4)
+ (u-t) (\epsilon_1\cdot\epsilon_4) (\epsilon_2\cdot\epsilon_3)
\nonumber \\
& & \quad
+ (\epsilon_1\cdot\epsilon_3)
       \Bigl[(\epsilon_2\cdot k_3) - (\epsilon_2\cdot k_4) \Bigr]
       \Bigl[(\epsilon_4\cdot k_2) - (\epsilon_4\cdot k_1) \Bigr]
\nonumber \\
& & \quad
+ (\epsilon_1\cdot\epsilon_4)
       \Bigl[(\epsilon_2\cdot k_3) - (\epsilon_2\cdot k_4) \Bigr]
       \Bigl[(\epsilon_3\cdot k_1) - (\epsilon_3\cdot k_2) \Bigr]
\nonumber \\
& & \quad
+ (\epsilon_2\cdot\epsilon_3)
       \Bigl[(\epsilon_1\cdot k_3) - (\epsilon_1\cdot k_4) \Bigr]
       \Bigl[(\epsilon_4\cdot k_1) - (\epsilon_4\cdot k_2) \Bigr]
\nonumber \\
& & \quad
- (\epsilon_2\cdot\epsilon_4)
       \Bigl[(\epsilon_1\cdot k_3) - (\epsilon_1\cdot k_4) \Bigr]
       \Bigl[(\epsilon_3\cdot k_1) - (\epsilon_3\cdot k_2) \Bigr]
\Bigr\} \,.
\label{b6}
\ea

\end{document}